\documentclass[fleqn,usenatbib]{mnras}

\usepackage[utf8]{inputenc}
\usepackage[T1]{fontenc}
\usepackage[toc,page]{appendix}

\usepackage[english]{babel}
\usepackage{float}
\usepackage{graphicx}

\usepackage{amsmath}                
\usepackage{amssymb}                

\usepackage{hyperref}
\usepackage{makecell}
\usepackage{calc}

\usepackage{multirow}
\usepackage{comment}
\usepackage{booktabs}

\usepackage{amsfonts}
\usepackage{amsmath}
\usepackage{amssymb}
\usepackage{nicefrac}
\usepackage{bm}

\usepackage{soul}
\usepackage{xspace}
\xspaceaddexceptions{]\}} 

\usepackage{braket}

\usepackage{enumerate}
\usepackage{enumitem}
\usepackage{array}
\usepackage{tabularx}

\usepackage{graphicx,array}
\usepackage[caption=false]{subfig}
\usepackage{epsfig}
\usepackage{longtable}

\usepackage{natbib}
\setcitestyle{round}
\usepackage[T1]{fontenc}
\usepackage{hyperref}

\title{Measuring satellite galaxy subhalo masses in redMaPPer clusters with UNIONS weak lensing data}

\author[R. Akhmetshyn et al.]{%
Roman Akhmetshyn,$^{1,2}$\thanks{roman.akhmetshyn@mail.mcgill.ca}
Jack Elvin-Poole,$^{1,3}$
Michael J. Hudson,$^{1,3,4}$
Isaac Cheng,$^{1}$ 
Thomas de Boer $^{5}$,
\newauthor Sébastien Fabbro$^{6}$,
 Sam Farrens$^{7}$,
Sacha Guerrini$^{7}$,
Axel Guinot$^{8}$,
Stephen Gwyn$^{9}$,
Hendrik Hildebrandt$^{10}$,
\newauthor Martin Kilbinger$^{7}$,
Charlie T. Mpetha$^{11}$, 
 Calum Murray$^{7}$, 
Ludovic Van Waerbeke$^{12}$,
Anna Wittje$^{13}$ 
\\
$^{1}$Department of Physics and Astronomy, University of Waterloo, Waterloo, Canada\\
$^{2}$Department of Physics, McGill University, Montréal, Canada\\
$^{3}$Waterloo Centre for Astrophysics, University of Waterloo, Waterloo, Canada\\
$^{4}$Perimeter Institute for Theoretical Physics, Waterloo, Canada\\
$^{5}$Institute for Astronomy, University of Hawaii, 2680 Woodlawn Drive, Honolulu, HI 96822, USA\\
$^{6}$NRC Herzberg Astronomy and Astrophysics, 5071 West Saanich Road, BC V9E2E7, Canada\\
$^{7}$Université Paris Cité, Université Paris-Saclay, CEA, CNRS, AIM, F-91191 Gif-sur-Yvette, France\\
$^{8}$Department of Physics, McWilliams Centre for Cosmology, Carnegie Mellon University, Pittsburgh, PA 15213, USA\\
$^{9}$Canadian Astronomy Data Centre, Herzberg Astronomy and Astrophysics, \\~National Research Council, 5071 West Saanich Rd, Victoria, BC V9E 2E7, Canada\\
$^{10}$Faculty of Physics and Astronomy, Ruhr University Bochum, Astronomical Institute (AIRUB),\\~German Centre for Cosmological Lensing, D-44780
Bochum, Germany\\
$^{11}$NASA Goddard Space Flight Centre, 8800 Greenbelt Rd, Greenbelt, MD 20771, USA\\
$^{12}$Department of Physics and Astronomy, University of British Columbia, 6224 Agricultural Road, BC V6T 1Z1, Vancouver, Canada\\ 
$^{13}$INFN Sezione di Ferrara, Via Saragat 1, I-44122 Ferrara, Italy
}

\date{Accepted XXX. Received YYY; in original form ZZZ}

\pubyear{2025}

\begin{document}
\label{firstpage}
\pagerange{\pageref{firstpage}--\pageref{lastpage}}
\maketitle

\begin{abstract}
In this work, we present the observed galaxy-galaxy lensing signal of satellite galaxies in rich clusters obtained from the Ultraviolet Near-Infrared Optical Northern Survey (UNIONS). Such observations are crucial for understanding dark matter halo physics and interaction dynamics in the cluster environment. Our goal is to investigate the tidal stripping of cluster subhalos. Theoretical predictions, supported by recent studies, suggest that as satellite galaxies fall into clusters, their dark matter halos are stripped and dispersed into the cluster's main halo, while their stellar mass remains relatively intact.
A robust method to probe this phenomenon is through statistical measurements of the excess surface mass density of satellite galaxies at different cluster-centric distances.  Our results reveal a significant lensing signal, with strong statistical power from the large number of lens-source pairs. A simple model used in the fitting process to constrain subhalo masses effectively reproduces the observed lensing signal at relevant satellite-centric distances. However, it lacks the complexity to accurately fit the excess surface density at larger distances, where the host cluster halo dominates. We measure the subhalo-to-stellar mass ratio as a function of cluster-centric radius and find a strong positive trend. These results confirm tidal stripping or other environmental processes that impact the relative masses of a satellite galaxy's stellar component and its dark matter halo.
\end{abstract}

\begin{keywords}
 gravitational lensing:weak -- galaxies:clusters:general -- galaxies:statistics -- dark matter
\end{keywords}



\section{Introduction}

In the standard cold dark matter model, structures grow hierarchically, with low-mass halos merging to form larger and more massive halos, such as galaxy clusters. 
However, as a small halo merges into a more massive halo, tidal interactions transfer mass from the infalling galaxies to the new host halo, leading to satellite galaxies partially loosing mass from their associated dark matter halos, now referred to as subhalos. Many theoretical predictions of a cluster's halo and subhalo interaction properties were developed using numerical simulations and semi-analytic modeling \citep[see e.g.][]{ghigna98, delucia04, gao04, contini13, bosch17, pullen14, du16}. They show that subhalos during their infall experience tidal stripping, tidal heating, and dynamical friction. Dark matter halos, being non-dissipative, experience more significant tidal disruption compared to the more tightly bound baryonic matter \citep{White1978}. Dynamical friction causes more massive subhalos to move more rapidly toward the centre, while tidal stripping primarily eliminates mass from the outer regions of subhalos situated near the host centre. This creates a hypothesis of evolution in galaxy clusters, that satellites will be largely stripped of their dark matter faster than their luminous matter. 
Simulations predict that the mass loss of infalling subhalos is inversely related to their distance from the centre of the halo (e.g. \cite{springel01}; \cite{delucia04}; \cite{gao04}; \cite{bower06}; \cite{xie15}; \cite{du16}). As a result, the halo mass to stellar mass ratio (HSMR, also referred to as the SHMR \citep{kumar2024, wang24}, or SMHM \citep{dvornik20}) for satellite galaxies is expected to decrease as they approach the host halo centre. Analysis of the HSMR of galaxies with similar stellar mass at different cluster-centric distances, and even including galaxies not in a cluster, is key in observing tidal interactions of dark matter.

We aim to probe the subhalo masses of satellite galaxies in clusters with gravitational lensing.
This phenomenon, which is well explained by general relativity, is a widely accepted technique for determining the overall projected mass of a galaxy, including its dark matter halo. 
Weak gravitational lensing, in particular, is frequently used to measure the average masses and structures of satellite galaxies and their halos (see \cite{li14}; \cite{li16}; \cite{sifon15}; \cite{sifon18}; \cite{uitern16};\cite{niemiec17}; \cite{dvornik20}; \cite{wang24}; \cite{kumar2024}).

Weak galaxy-galaxy lensing (where both source and lens are galaxies) is an intrinsically statistical observational measurement. Unlike strong lensing, the distortion induced in each background galaxy is much smaller than the typical galaxy ellipticity.  Measuring the weak lensing signal
around satellite galaxies is challenging due to the small relative contribution of the satellite to the lensing signal produced by the galaxy cluster. 
Nevertheless, it is possible to identify the collective average tangential shear produced by a large group of lenses by combining measurements from numerous individual lensing signals. This way, the statistical noise can be averaged out and the mean lensing of satellite galaxies can be measured. The lensing amplitude can then be used to characterize subhalos via excess surface density (more in the next section). The subhalo mass of cluster satellites has been measured previously in \cite{li14}, \cite{sifon15},  \cite{li16}, \cite{niemiec17}, \cite{sifon18}, \cite{wang24}, and \cite{kumar2024}, using source ellipticities from  CFHT Stripe-82 survey, Kilo-Degree Survey, DECaLS DR8, CFHT  Lensing Survey, and Hyper Suprime-Cam survey. 

In this paper, we use background galaxies with high-quality shape measurements from the Ultraviolet Near Infrared Optical Northern Survey (UNIONS) as described in \cite{guinot22}, and satellite galaxies from the redMaPPer SDSS DR8 galaxy cluster catalog \citep{rykoff16}. The uniqueness of our work lies in the amount of data we have for the excess surface density measurement. The combined number of selected satellites is almost 330000, while the number of background galaxies from the survey that overlaps redMaPPer clusters is roughly 4.5 million. These large numbers contribute to smaller error bars in obtained density measurements, which in turn can help us to better fit the density model and constrain parameters.

In our models and data we adopt a flat $\Lambda$CDM cosmology, with the Hubble parameter $H_0=70$ km s$^{-1}$ Mpc$^{-1}$, density of matter $\Omega_{\rm M}=0.3$, baryon density $\Omega_{\rm B}=0.049$,  both are in units of the critical density, power spectrum normalization $\sigma_{8}=0.81$, tilt of the primordial power spectrum $n_{\rm s}=0.95$. All quantities are displayed in proper units, where the dependence on $h$ is dropped, unless written otherwise. In places where data had to be converted from different cosmologies, an appropriate conversion was applied \citep[see e.g.][]{croton13}.

The layout of this paper is as follows. Section \ref{sec:model} describes how we compute the combined lensing signal, including terms representing a subhalo, an offset host halo, and the stellar mass. Section \ref{sec:data} contains information on the lens and source catalogs used. In Section \ref{sec:compute}, we summarize the excess surface density computation process. Finally, Section \ref{sec:results} through \ref{sec:concl} describe our fitting process, display key results, and discuss them. 

\section{Theoretical model}
\label{sec:model}

Excess surface mass density $\Delta\Sigma (R)$ is a function
of the distance $(R)$ to the gravitational lens centre, and is related to
the tangential shear $\gamma_{\rm t}$:
\begin{align}
    \Delta\Sigma(R)=\Sigma_{\rm crit}\gamma_{\rm t}(R)=\overline{\Sigma}(<R)-\overline{\Sigma}(R) \,.
    \label{eqn:surface_density}
\end{align}
Here $\overline{\Sigma}(R)$ is the mean surface density at distance $R$; $\overline{\Sigma}(<R)$ is the mean surface density enclosed within radius $R$; $\Sigma_{\rm crit}$ is critical surface density:
\begin{align}
    \Sigma_{\rm crit}(z_{\rm l},z_{\rm s})=\frac{c^2}{4\pi G}\frac{D_A(z_{\rm s})}{D_A(z_{\rm l})D_A(z_{\rm l},z_{\rm s})}
    \label{eqn:sigma_crit}
\end{align}
The function above characterizes the relative amplitude of light deflection and depends on the angular diameter distances $D_{\rm A}$ to the source ($z_{\rm s}$) and lens ($z_{\rm l}$) redshifts.

The lensing signal measured around satellites in galaxy clusters contains contributions from the host halo, subhalos, and satellites' baryonic matter. We thus use combined excess surface mass density profiles in our model:
\begin{align}
    \Delta\Sigma(R)= A\cdot\Delta\Sigma_{\rm host} + \Delta\Sigma_{\rm sub}
+ \Delta\Sigma_{\rm *}
    \label{eq:3}
\end{align}
Note that $R$ is a satellite-centric distance, which means we must calculate the host halo profile offset from the cluster centre. Following the procedure of \cite{li16}, we introduce a host halo rescaling factor, $A$, which captures inaccuracies in the halo mass estimation and corrects for model-data discrepancies. If the mass-richness relation and concentration of the host are known accurately, the best-fit $A$ should be consistent with unity. 

Just like in \cite{li14}, \cite{li16}, \cite{niemiec17} and \cite{wang24}, a contribution from the stellar
component of the satellite galaxy ($\Delta\Sigma_{\rm *}(R)$) is added to the excess density model and is calculated as for a point mass:
\begin{align}
    \Delta\Sigma_{\rm *}(R)=\frac{\langle M_{\ast} \rangle}{\pi R^2}
\end{align}
where $\langle M_{\ast} \rangle$ is the average stellar mass of a satellite. We include it in our final fit, as it can help to better constrain subhalo contribution at very close satellite-centric distances.

\cite{niemiec17} included a two-halo term, $\Delta\Sigma_{\rm 2h}(R)$, produced by the neighbouring halos. However, as found in that paper, it has little influence up to 1.8 $h^{-1}$Mpc, 2.6 Mpc in our cosmological framework. 
We do not include the two-halo term in our model, because our analysis focuses on $R < 0.6$ Mpc, where the subhalo term is dominant. Nevertheless, we believe, that for galaxies in the out-most cluster-centric distance bin (0.6--0.9 $h^{-1}$Mpc), the two-halo term could influence the slope of the signal at larger R. The topic of two-halo term is discussed in Section \ref{sec:res_disc}.

\subsection{Host halo profile}
Density profiles of dark matter halos can be described by
the Navarro–Frenk–White 
\cite[NFW;][]{navarro95}
profile:
\begin{align*}
    \rho (r) = \frac{\rho_{\rm s}}{\frac{r}{r_{\rm s}}\left(1+\frac{r}{r_{\rm s}}\right)^2} \,,
\end{align*}
where $r_{\rm s}$ is a scale radius, and $\rho_{\rm s}$ is central density. These parameters can also be defined using spherical overdensity mass ($M_{\rm halo}$) and concentration ($c$). We assume the mass-concentration relation from \cite{duffy08}.

To model the $\Delta\Sigma (R)$ contribution from the host profile, we use a statistical approach, accounting for the offset between the host and subhalo. Specifically, we use Monte-Carlo techniques to simulate randomly placed $\times10^{10}$M$_{\odot}$ point masses that follow a cluster-centric 2D NFW profile for a given halo mass, cluster photometric redshift, and concentration. The photometric redshifts are taken from the redMaPPer cluster catalog \citep{rykoff16} compiled from Dark Energy Survey
 (DES, \cite{DES}) data. The number of simulated points depends on the cluster mass. The profiles were computed using the \texttt{Colossus} package in Python (\cite{diemer18}). Note, this differs from the analytical approach used in other works \citep{li14, sifon15, sifon18, uitern16, niemiec17, li16, wang24}. In principle, both Monte-Carlo and analytical approach produce the same results, and there is no apparent advantage in one or the other. However, the former is is easier to implement computationally, especially in the case of off-centring of the cluster centre, that would otherwise require double integrals (see Section \ref{off-center}).

Halo masses were derived from the observed cluster richness from the redMaPPer catalog \citep{rykoff16} using the following empirical relation from \cite{rykoff12} (Equation B4):
\begin{align}
    M_{\rm 200m}=e^{1.72}\left(\frac{\lambda}{60}\right)^{1.08}\cdot h_{70}^{-1}10^{14} \ M_{\odot},
\end{align}
where $\lambda$ refers to the richness parameter of a cluster, $h_{70} =H_0/70$ km s$^{-1}$ Mpc$^{-1}$ is the dimensionless Hubble constant, and is numerically assumed to be 1 for this analysis. Halo masses throughout this work are defined by the enclosed density of 200 times the mean density of the Universe. 

\begin{figure}
    \centering
    \includegraphics[width=0.45\textwidth]{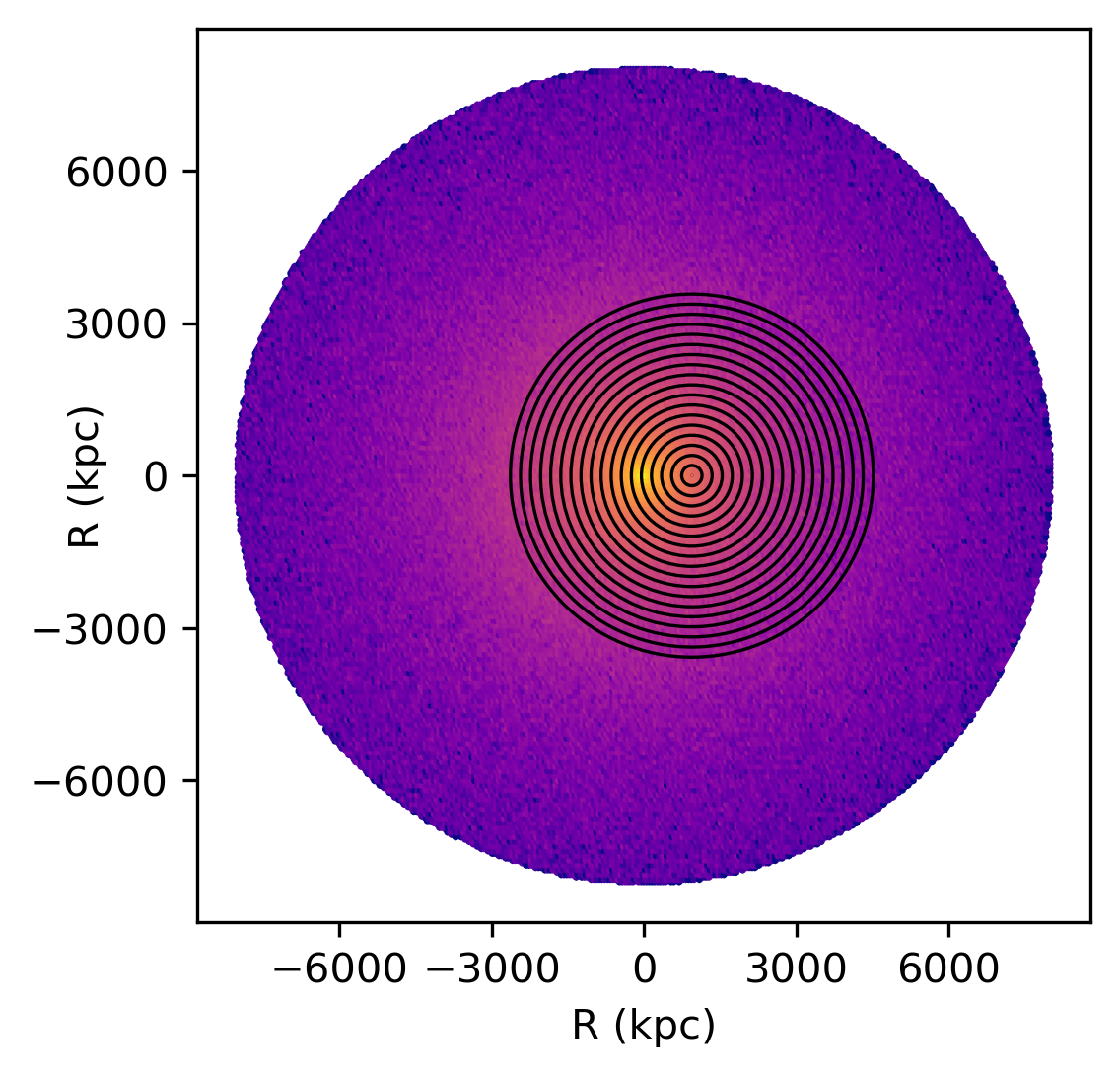}
    \caption{Demonstration of the offset halo profile simulation. The colour gradient represents Monte-Carlo point masses that follow the enclosed mass profile with a centre (0,0) at the cluster centre. Black concentric rings, centred at a given cluster-satellite distance ($r$), are used to calculate $\Sigma(R)$ and $\Sigma(<R)$, where $R$ is satellite-centric distance. The width of each ring for $\Sigma(R)$ calculation is 1.4 kpc, and the distance between rings is 28 kpc. Compared to the scales of the simulation, the width is nearly infinitesimal. These parameters were increased in this demonstration figure.}
    \label{halo_model}
\end{figure}

We compute the surface density $\Sigma_{\rm host}$ with Monte-Carlo point masses within a given satellite-centric ring and use Equation \ref{eqn:surface_density} to convert to $\Delta \Sigma_{\rm host} (R)$. This procedure is illustrated in Figure \ref{halo_model}. The final results are obtained by averaging excess surface density calculations for all satellite galaxies and their respective clusters that are in a given cluster-centric ($r_p$) distance bin. The total number of offset profiles  in each bin (N$_{\rm sat}$) that are averaged is referred in Table \ref{tab:param_fits}. As in \cite{li14}, we separate our lenses into 3 cluster-centric bins: 0.1--0.3, 0.3--0.6, and 0.6--0.9 h$^{-1}$Mpc. 
Which, in our case, correspond to 0.14--0.43, 0.43--0.86, and 0.86--1.29 Mpc bins. Each bin has 26059, 21206, and 5274 unique clusters respectively, but only 6000, 4854, and 1152 overlap with the UNIONS background galaxies (galaxies within R$_{\rm200m}$ of a cluster).

\subsection{Uncertainty in the cluster-centric distance}
\label{off-center}
Our initial analysis assigned the cluster-satellite distance ($R$) to be the distance to the most likely Brightest Central Galaxy (BCG) reported in the redMaPPer catalog \citep{rykoff16}. This choice produced a sharp dip into negative regions of $\Delta \Sigma_{\rm host} (R)$. As illustrated later in our results, the observed trough in the lensing signal is more smeared and rarely goes negative. We thus explore a possibility, that the reported BCG position may be incorrectly defined, or offset  from the actual centre of the corresponding dark matter halo, or gravitational potential minimum of the cluster. The resulting cluster-satellite distance would also shift from the reported value, and consecutively, the averaged offset halo signal should be more smoothed.
The idea of a mis-centred BCG was explored in previous studies \citep{Johnston07, kumar2024}. We follow their approach, which assumes that the position of the ``true''\ BCG, follows a radially-uniform Rayleigh distribution around the cataloged BCG position:
\begin{figure}
    \centering
    \includegraphics[width=0.48\textwidth]{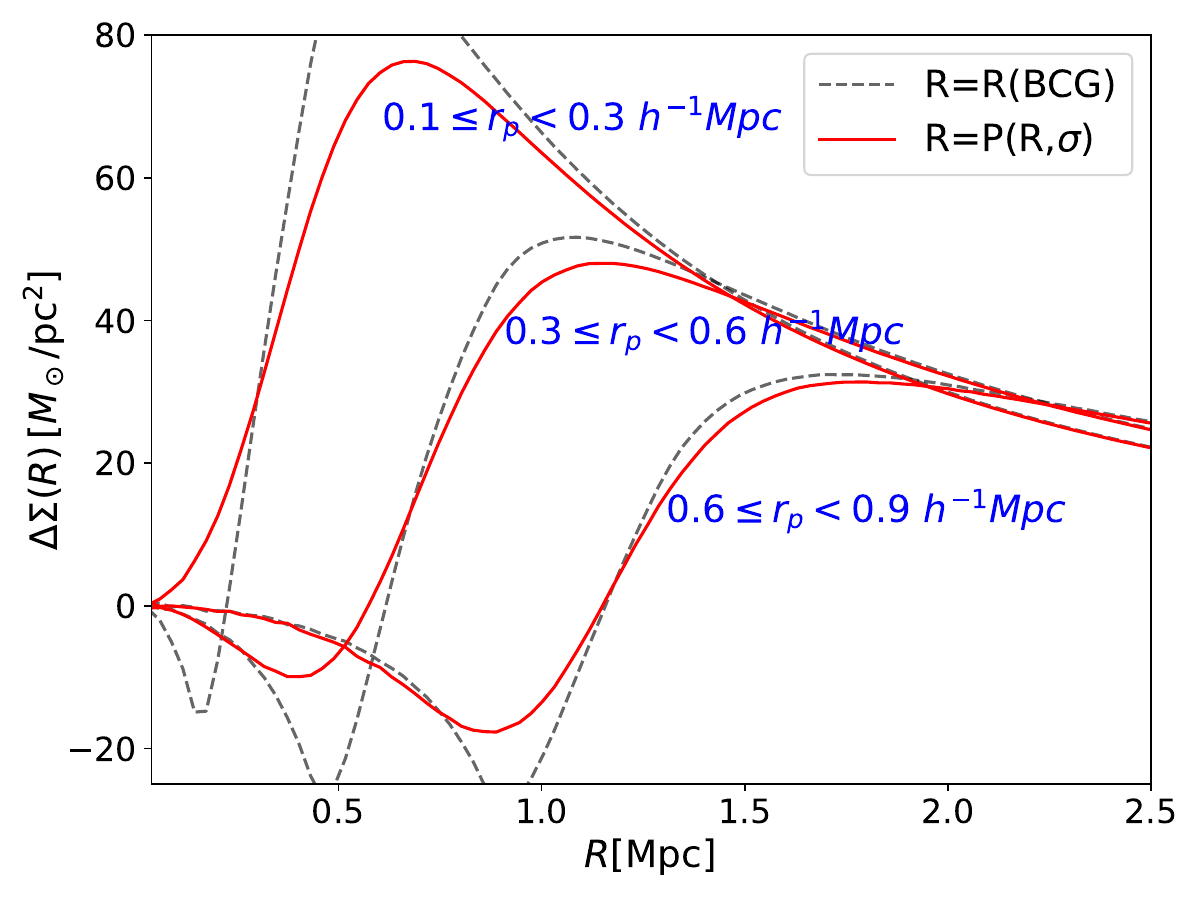}
    \caption{Offset halo term for each cluster-centric distance bin, indicated by the blue text beside, generated using 2 methods. Black dashed lines show the halo profile computed with absolute cluster-satellite distances reported by the redMaPPer catalog. The solid red line is for a Rayleigh-distributed BCG position with a $\sigma=143$ kpc applied.}
    \label{fig:hostmodel}
\end{figure}
\begin{equation}
    P(R_{\text{off}}) = \frac{R_{\text{off}}}{\sigma^2}e^{-\frac{R_{\text{off}}}{2\sigma^2}}
    \label{eqn:rayleigh}
\end{equation}
Here, $R_{\text{off}}$ is the offset distance between the catalog's BCG and true centre, $\sigma$ is the scale parameter, and also the mode of the distribution. In our fitting process, it is one of the free parameters. The ``new'' cluster-centric distance is then found using the cosine law:
\begin{equation}
    r_{\text{new}} = \sqrt{r^2_{\text{cat}} + R_{\text{off}}^2 - 2R_{\text{cat}} R_{\text{off}} \cos{\theta}}, \quad \theta \sim \mathcal{U}(0, 2\pi)
\end{equation}
The offset cluster halos produced with applied random BCG offset and without can be compared in Figure \ref{fig:hostmodel}. While we filter redMaPPer lenses into respective $r_p$ bins using absolute cataloged distances, the assigned distance to a ``new'' BCG may exceed that bin. This creates a smeared averaged profile and results in a wider and shallower trough. This morphology is supported by the observed signal. We create a number of offset profiles with a range of $\sigma$ values, and later during the fitting process, we compute an instantaneous interpolated profile for a given $\sigma$.

\begin{figure*}
    \centering
    \includegraphics[width=1\textwidth]{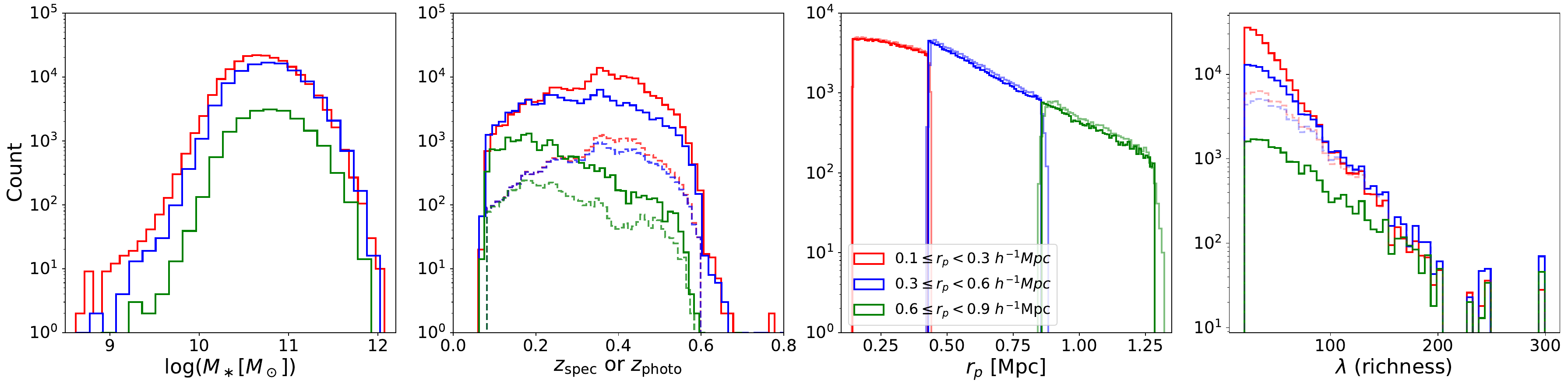}
    \caption{From left to right: histograms of lensing galaxy stellar masses ($\log M_\text{*}$), satellite and cluster redshifts ($z$), cluster-centric distances ($r_p$), and richness ($\lambda$) values associated with each satellite galaxy used to compute the offset halo term for the 3 bins. These bins are in the same units and cosmological dependency as in  \protect\cite{rykoff16},  \protect\cite{li16}, and  \protect\cite{wang24} for convenience.
    Dashed histograms on the second plot indicate photometric redshifts of host halos. The dashed histograms on the rightmost plot show richness distributions if only satellites that are in common halos are considered. Cluster-centric distances in faint lines overlap because we used distances for the likeliest BCG, instead of a weighted mean to account for some uncertainty. Mean values for $\log M_\text{*}$, ~$z$, and $\lambda$ are provided in Table \ref{tab:param_fits}.}
    \label{fig:lens_stat}
\end{figure*}

\subsection{Subhalo profile}

By projecting the NFW 3D density profile for a subhalo $\rho(r)$ along the line of sight, we can compute its surface density $\Sigma_{\rm sub} (R)$ profile:
\begin{align}
    \Sigma_{\rm sub}(R)=2\int_R^{\infty} \frac{r\rho(r)}{\sqrt{r^2-R^2}}dr
\end{align}
which can be converted to a $\Delta \Sigma$ profile using Equation \ref{eqn:surface_density}. We used the Python module \texttt{profiley} to compute the NFW profile of the subhalo term \citep{madhavacheril20}.

\section{Observational data}
\label{sec:data}

\subsection{Lens galaxies}
In this study, we use the redMaPPer SDSS DR8 cluster and member catalogs for the selection of gravitational lenses. The redMaPPer algorithm is used to identify galaxy clusters through a probabilistic percolation approach \citep{rykoff14}, primarily relying on the red sequence of galaxies of the SDSS Data Release 8 (DR8, \cite{aihara11}). The cluster catalog covers 10134 deg$^2$ of the sky and contains 26111 galaxy clusters that are in  redshift range of $0.08<z<0.6$ and richness of $>$20 \citep{rykoff16}. The members catalog contains over 1.7 million galaxies that are matched with their respective clusters using their various properties (color, \textit{ugri} magnitudes, and spatial distribution).

We apply a membership probability threshold of 0.8 to filter the member catalog. This threshold is supported in \cite{niemiec17},  \cite{li16}, and \cite{wang24}, and ensures that we reduce contamination from field galaxies. This filter, in addition to our limitations on distance from cluster centre (0.1$ < R_p\leq 0.9$ h$^{-1}$Mpc), reduces the number of lenses to roughly 330,000. However, the majority of those members do not have SDSS spectroscopic redshifts. For members with unknown spectroscopic redshifts, we assign their cluster's photometric redshift, which is estimated based on the redshift distribution of its member galaxies. This practice is supported by our analysis, which indicates that the difference between the cluster's photometric redshift and the spectroscopic member redshifts is consistently small, especially for high-probability cluster members. This approach allows us to include a larger number of lenses in our analysis, while including minor uncertainty in the lens redshifts. More information on members catalog filtering is in Appendix \ref{sec:redshift}.

\subsection{Source galaxies}
UNIONS is a collaboration between the Canada-France Imaging Survey (CFIS), the Panoramic Survey And Rapid Response System \citep[Pan-STARRS;][]{chambers16}, WISHES (Wide Imaging with Subaru HSC of the Euclid Sky), and WHIGS (Waterloo-Hawaii Institute
for Astronomy g-band Survey). It aims to provide the largest (6250 deg$^2$) multi-band optical photometric survey of the northern hemisphere. The survey area overlaps with wide spectroscopic surveys such as SDSS’s Baryon Oscillation Spectroscopic Survey \citep[SDSS-BOSS;][]{eisenstein2011sdss},
extended Baryon Oscillation Spectroscopic Survey (eBOSS;
\cite{hutchinson16}), and the Dark Energy Spectroscopic
Instrument (DESI) survey (\cite{desi}). Full details of the UNION Survey can be found in \cite{gwyn2025}.

Background galaxies and their ellipticity parameters were computed using ShapePipe (v1.3), an open-source Python weak-lensing pipeline that produces accurate galaxy shape measurement, shear calibration, extensive validation tests on the point spread function and on the galaxy shapes (see \cite{guinot22} for details).

Lenses and source catalogs probe different parts of the sky, yet they have a substantial mutual coverage, which is a huge statistical benefit. According to our rough estimates, the total number of sources within R$_{\rm200m}$ from BCGs is 4,454,600. The average number of sources per cluster is 461, and the total number of clusters overlapping UNIONS data is 6012  out of 26111. The total overlapping area of lenses and sources from selected cluster-centric bins is roughly 2310 deg$^2$.

\subsection{Stellar mass of the lenses}

 Satellite stellar masses were retrieved from an analysis of the Dark Energy Spectroscopic Instrument  Legacy Imaging Surveys by \cite{zou19}.
 We cross-matched the their catalog with the redMaPPer catalog and found 96.8\% matches. Statistics on lensing galaxies' masses, redshifts, and cluster-centric distances are displayed in Figure \ref{fig:lens_stat}.

\section{Excess surface density computation}
\label{sec:compute}

In this paper, we use the open-source Python package \texttt{dsigma} \citep{lange22} for measuring galaxy-galaxy lensing. The core computations are written in C, thus making it computationally fast. Additionally, \texttt{dsigma} provides support for estimating covariances with jackknife resampling.

A good first-order estimate of $\Delta\Sigma$ can be calculated using critical surface density $\Sigma_{\rm crit}$ (see Equation \ref{eqn:sigma_crit}):
\begin{align}
     \Delta\Sigma=\frac{\sum_{ls}w_{ls}\Sigma_{\rm crit}(z_{\rm l},z_{\rm s}) e_t}{\sum_{ls}w_{ls}}
     \label{dsigma}
\end{align}
where
\begin{align}
     w_{ls}=\frac{w_s}{\Sigma^2_{\rm crit}(z_{\rm l},z_{\rm s})}
\end{align}
Here $w_s$ is a source weight that takes into account uncertainties in shape measurements, $e_t$ is the tangential ellipticities of the sources. Calculation of $\Delta\Sigma$ at different satellite-centric distances requires summation over all suitable lens-source pairs. The \texttt{dsigma} module allows us to efficiently find and compute such pairs in the case of a large number of sources and lenses. We also use this algorithm to apply lens selection bias correction, redshift correction using the ShapePipe calibration catalog. To correct for any over-abundance of sources close to lenses, the algorithm multiplies the raw lensing signal by the radially dependent boost factor:

\begin{equation}
    b =
        \frac{\sum_{ls} w_{\mathrm{sys}, l} w_{ls}}{\sum_{rs}
              w_{\mathrm{sys}, l} w_{ls}} \, ,
\end{equation}
Where $w_{\mathrm{sys}, l}$ refers to systematic weights applied during the lens selection bias correction, and the summation in the denominator is done over a random set of lenses with a similar redshift as the actual lenses.

Since, at this time, we do not have photometric redshift estimates available for individual UNIONS galaxies, we follow the procedure of \cite{cheng2025} and \cite{liUNIONS} to correct for the broad $n(z)$ of the UNIONS galaxies. We compute the mean $\Sigma_{\rm crit}(z_{\rm l})$ over the source catalog using a catalog of known spectroscopic redshifts detailed in \cite[appendix A]{liUNIONS} using,

\begin{equation}
    \langle \Sigma^{-1}_{\rm crit}(z_{\rm l}) \rangle = \int \Sigma^{-1}_{\rm crit}(z_{\rm l}, z_{\rm s})n(z_{\rm s}) dz_{\rm s}. 
\end{equation}

The estimator in Eqn.\ref{dsigma} then becomes, 

\begin{equation}
     \Delta\Sigma=\frac{\sum_{ls}w_{ls}\langle\Sigma_{\rm crit}^{-1}(z_{\rm l})\rangle ^{-1}e_t}{\sum_{ls}w_{ls}}
\end{equation}

We calculate the lensing signal at 14 satellite-centric bins centred at [0.101, 0.175, 0.248, 0.320, 0.423, 0.567, 0.711, 0.854, 1.025, 1.226, 1.429, 1.634, 1.837, 2.040] Mpc. The data points were selected to be more concentrated at low $R$, where, in theory, the combined lensing signal is largely contributed by the subhalo. This will allow us to better constrain subhalo mass ($M_{sub}$). 

\section{Fitting Method}
\label{sec:results}

The lensing signal from each of 3 cluster-centric distance bins ($0.1\leq r_p<0.3$, $0.3\leq r_p<0.6$, $0.6\leq r_p<0.9$ $h^{-1}$Mpc) is fitted individually with the combined model (eq.\ref{eq:3}) using the Monte-Carlo-Markov-Chain (MCMC) method from the \texttt{emcee} package \citep{emcee} with Metropolis–Hastings moves. For our final result, we have 3 free parameters: $M_{\rm sub}$, the subhalo mass for HSMR analysis, $A$, the cluster halo mass rescaling factor, and $\sigma$, the scale parameter of the Rayleigh distribution for the BCG offset distance. The rescaling factor is not fixed across 3 $r_{\rm p}$ bins: as seen from Figure \ref{fig:lens_stat}, each bin has a slightly different distribution of host halos. For example, on average, the outermost bin contains less massive halos, resulting in a smaller amplitude of the averaged offset halo profile. Parameter $A$ adjusts for this offset and accounts for the uncertainty in richness, as well as the richness to mass estimation discrepancy. Similarly, $\sigma$ is not fixed to account for this bias. We use 50 walkers with 20000 steps, and burn-in at 10000 steps, step size follows a normal distribution with mean 0 and standard deviation of 0.01, 0.01, and 1.0 for the respective parameters. A flat prior range is adopted for each free parameter:
\begin{itemize}
    \item $10 \leq \log \left( {M_{\rm sub}} / \textup{M}_\odot \right) \leq 13$
    \item $0 \leq \textup{A} \leq 1.5$
    \item $0 \leq \left( \sigma / \text{kpc}\right) \leq 500$
\end{itemize}
The log likelihood function for the MCMC is proportional to the $\chi^2$ value, which is calculated as:
\begin{align}
    \chi^2&=(\mathbb{D}-\mathbb{M})^T\cdot \mathbb{C}^{-1}\cdot(\mathbb{D}-\mathbb{M}) \text{ ,}\\
  \text{thus }  \ln L &\propto -\frac{1}{2} \chi^2
\end{align}
where $\mathbb{D}$ is the observed data vector, $\mathbb{M}$ is the model we fit and $\mathbb{C}$ is the covariance matrix. We use jackknife resampling to compute the covariance matrix.  Bias in $\mathbb{C}^{-1}$  leads to an underestimation of the size of confidence regions by making the log-likelihood function steeper (\cite{hartlap07}). That is why we multiply our covariance matrix by the Hartlap factor F$_{\rm H}$ before calculating the inverse of $\mathbb{C}$:
\begin{align}
    F_{\rm H}&=\left( \frac{n-p-2}{n -1} \right)^{-1} \text{ ,}\\
    \text{thus } \mathbb{C}&=\mathbb{C}\cdot F_{\rm H}
\end{align}
This bias depends on the ratio of the number of data points $p$ to the number of independent observations $n$. In our case, it is the number of jackknife regions, which is $n=100$.

\begin{figure}
    \centering
    \includegraphics[width=0.46\textwidth]{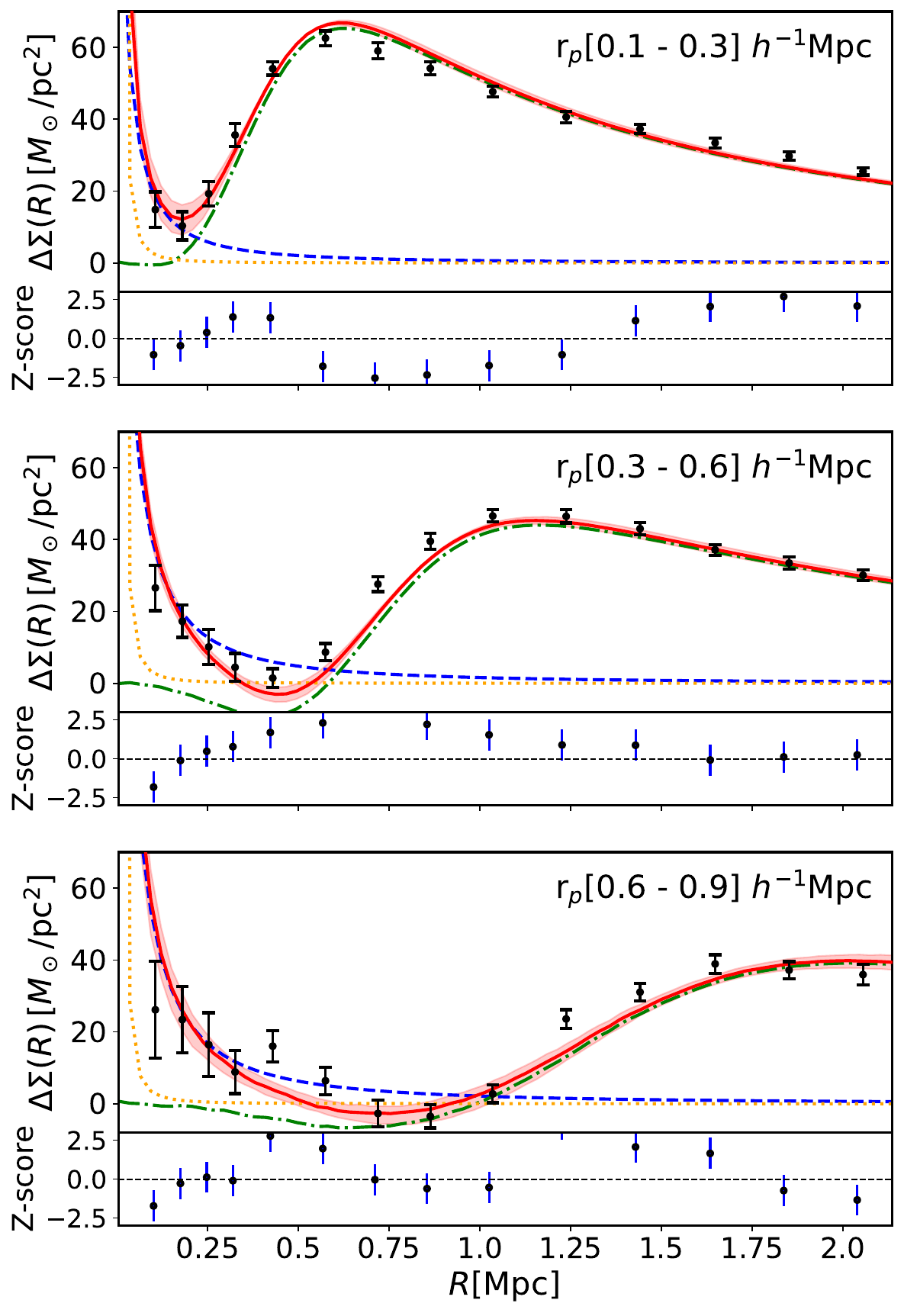}
    \caption{Satellite-centric lensing signal and best fit result for each cluster-centric $r_p$ bin. The error bars are estimated using jackknifing with 100 regions. The red solid line is the combined signal that is a sum of 3 terms in our model: stellar (orange dashed), subhalo (blue dashed), and offset halo (green dashed). Stellar component is constant; offset halo is an interpolation of precomputed models for a given $\sigma$ parameter, and is multiplied by a free parameter A; subhalo term is constructed by estimating subhalo mass. The translucent red area shows the range between the 16th and 84th percentiles of MCMC samples.  Below each subplot is the corresponding residual graph in units of standard deviation.}
    \label{results}
\end{figure}

\section{Results and Discussion}
\label{sec:res_disc}

In Figure \ref{results} we show the measured lensing signal at different cluster-centric bins, and the best-fit model with a solid red line. Our overall calculated signal morphology is in agreement with calculations achieved by \cite{kumar2024}, \cite{wang24}, \cite{niemiec17}, and \cite{li16}, with some deviations in the amplitude. The adopted final model generally follows the observed curve in excess surface density, however, some regions remain poorly explained. The most noticeable is the rising slope to the right from the trough: for each bin, we underestimate the lensing signal. We suspect that a more quantifiable approach is needed for simulating uncertainties in cluster-satellite distance estimations. Perhaps the offset BCG distance follows a different distribution, or, more likely, is not uniform in angle space. The other noticeable discrepancy is the slope of the fit for the \hbox{$0.1\leq R_p<0.3$} and $0.6\leq R_p<0.9$ $h^{-1}$Mpc bins at distances after the main peak. For the closest bin, it could be the effect of other neighbour subhalos that are not taken into account. For the furthest bin, a two-halo term, used in \cite{niemiec17}, could improve signal estimation, as its contribution increases with distance. The last noticeable area is where 2 data points that exceed the model near the 0.5 Mpc region in the $0.6\leq R_p<0.9$ Mpc bin. This was also observed in \cite{li16}, and appears to be a statistically significant part of the signal measurement. We currently do not have a viable explanation for such behavior. Overall, the error bars for the last bin are larger as the number of satellite lensing galaxies is smaller. There could also be a higher chance of member mismatch at those distances. \cite{Sunayama19} explored  halo assembly bias in the redMaPPer catalog and concluded that the membership probability
cannot really distinguish between the interlopers and real members, even at P$_{\rm mem}>0.8$. This implies that our lensing signal may be contaminated by field galaxies, which could explain some features discussed above.

\begin{figure}
    \centering
    \includegraphics[width=0.99\linewidth]{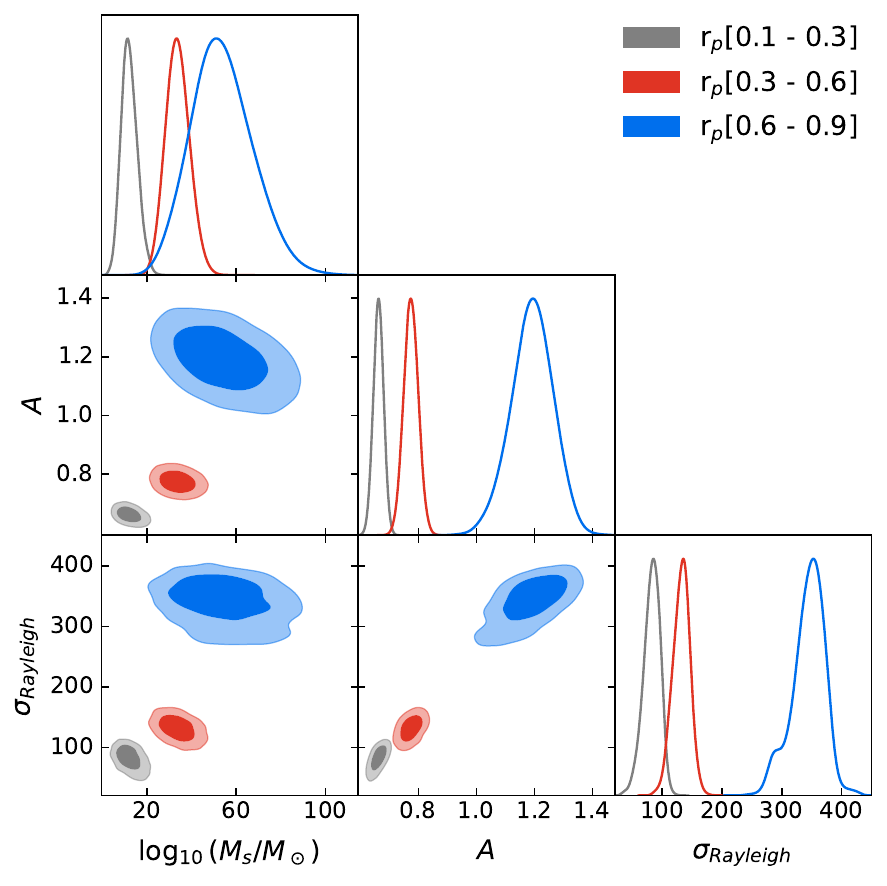}
    \caption{Corner plot of the best-fit model for each bin. Grey for the closest cluster-centric bin, red for the intermediate, and blue for the furthest. Solid contours around scattered samples indicate 68\% and 95\% of samples, and histograms are marginalized posterior distribution for each fitted parameter (satellite's subhalo mass devided by respective stellar mass (HSMR), offset halo scale factor, and Rayleigh distribution $\sigma$ parameter).}
    \label{fig:final_corner}
\end{figure}

The combined corner plot in Figure \ref{fig:final_corner} shows the MCMC posterior distribution for marginalized model parameters that best describe the observed signal for all $r_p$ bins. Posterior samples of the subhalo masses were divided by average stellar masses to show the posterior distribution of the HSMR. Reported uncertainties are based on the 16th and 84th percentiles of the samples. Although all MCMC chains converged, the $\sigma$ parameter for the $0.6\leq R_p<0.9$ Mpc bin displays a non-Gaussian posterior distribution. This is likely caused by the scatter in the lensing signal. If we remove the datum at 1.25 Mpc, the $\sigma$ posterior is closer to the normal distribution.

The host halo scale factor $A$ is correlated with the cluster-centric bins.
This may be explained by each bin having a different sample of host halos and a specific way of computing the offset halo model. This correlation could be dampened if we applied richness parameter of each cluster as weights in the computation of an averaged offset halo profile. In previously mentioned \cite{li16} and \cite{wang24}, the halo normalization factor should not depend on satellite bin samples and/or is close to unity. This factor also correlates with the Rayleigh distribution $\sigma$ scale parameter, as it also increases with $r_p$ bins. This may suggest that the uncertainty in BCG-satellite distance is higher for host halo sample of the furthest bin, and the estimate on $\sigma$ means that on average, the ``real'' BCG is located further away from the redMaPPer reported BCG position. 

We theorized that a way to reduce the difference in these two parameters across $r_p$ bins would be to restrict the lens selection to galaxies residing only in common host halos. After applying this criterion, the sample contained 134669 lenses in total, fewer than in the initial sample (see Table \ref{lenstable} for details). Following the same procedure to compute the lensing signal and the offset halo profiles, we find that, first, the amplitude of the halo term has increased for the first two closest $r_p$ bins. This is expected, as the common-halo cut removed a significant contribution from low-mass (low-richness) halos (see Figure \ref{fig:lens_stat}). Second, the scaling factor $A$ moved closer to unity: $0.80^{+0.03}_{-0.03}$ and $0.88^{+0.04}_{-0.04}$ for the \hbox{$0.1\leq R_p<0.3$} and $0.3\leq R_p<0.6$ $h^{-1}$Mpc bins respectively, while remaining unchanged for the furthest bin. We believe that introducing the number of galaxies within a halo as weights during computation of the excess surface density and the offset halo profile could bring the scaling factor even closer to unity.
However, the $\sigma$ parameter diverged in a less expected way: $69^{+22}_{-29}$, $143^{+36}_{-23}$, and $321\pm18$, in units of kpc, from the closest to the furthest bin, respectively, although these values remain consistent with our reported results within the uncertainties. The correlated mismatch between $A$ and $\sigma$ could therefore be attributed to other effects, which may have a stronger impact on the physical interpretation of the results and warrant further investigation. 

\begin{table*}  
\centering
\caption{Our best-fit estimations of the median satellite halo mass ($\log\langle M_{\rm sub}\rangle$), halo mass rescaling factor ($A$), HSMR number ($M_{\rm sub}$/$M_{*}$), Rayleigh distribution scale parameter($\sigma$) with errors based on the 16 and 84 percentile from the posterior
 distributions, as well as reduced $\chi^2$ (11 degrees of freedom) number for each cluster-centric bin. Known parameters such as mean stellar mass ($\log\langle M_{*}\rangle$) from \protect\cite{zou19}, mean redshifts ($\langle z_{\rm sat}\rangle$), mean host cluster richness ($\langle \lambda\rangle$), mean membership probabilities ($\langle \rm P_{\rm mem}\rangle$) of lens samples are also provided. N$_{\rm sat}$ refers to the sample size of lensing satellites for each bin. P$_{\rm mem}$ refers to the median cluster membership probability of the lens sample.}
\begin{tabular}{@{}lllcllllccc@{}}
\toprule
$r_p$ [Mpc] & log$\langle M_{*}\rangle$ & log$\langle M_{\rm sub}\rangle$ & A  &$\sigma$
[kpc]& $M_{\rm sub}$/$M_{*}$  &$\chi^2_{\text{red}}$& $\langle z_{\rm sat}\rangle$ & N$_{\rm sat}$&$\langle\lambda\rangle$ &P$_{\rm mem}$\\ \midrule
0.143--0.43 & $10.86$ &$12.02^{+0.12}_{-0.14}$& $0.66^{+0.02}_{-0.02}$&$83.51^{+12.89}_{-14.87}$&$11.99^{+3.90}_{-3.34}$&3.59& 0.348 & 204000  &38.77 &0.91\\
0.43--0.86 & $10.92$ &$12.45^{+0.07}_{-0.07}$& $0.77^{+0.03}_{-0.03}$&$132.71^{+13.26}_{-15.47}$&$33.68^{+5.53}_{-5.17}$&2.77& 0.312 & 106586  &40.73 &0.86\\
0.86--1.23 & $10.94$ &$12.58^{+0.11}_{-0.12}$& $1.19^{+0.07}_{-0.07}$&$347.70^{+23.35}_{-29.68}$&$53.02^{+14.73}_{-12.98}$&3.39& 0.229 & 18865  &48.99 &0.85\\ \bottomrule
\end{tabular}

\label{tab:param_fits}
\end{table*}

\begin{figure}
    \centering
    \includegraphics[width=0.46\textwidth]{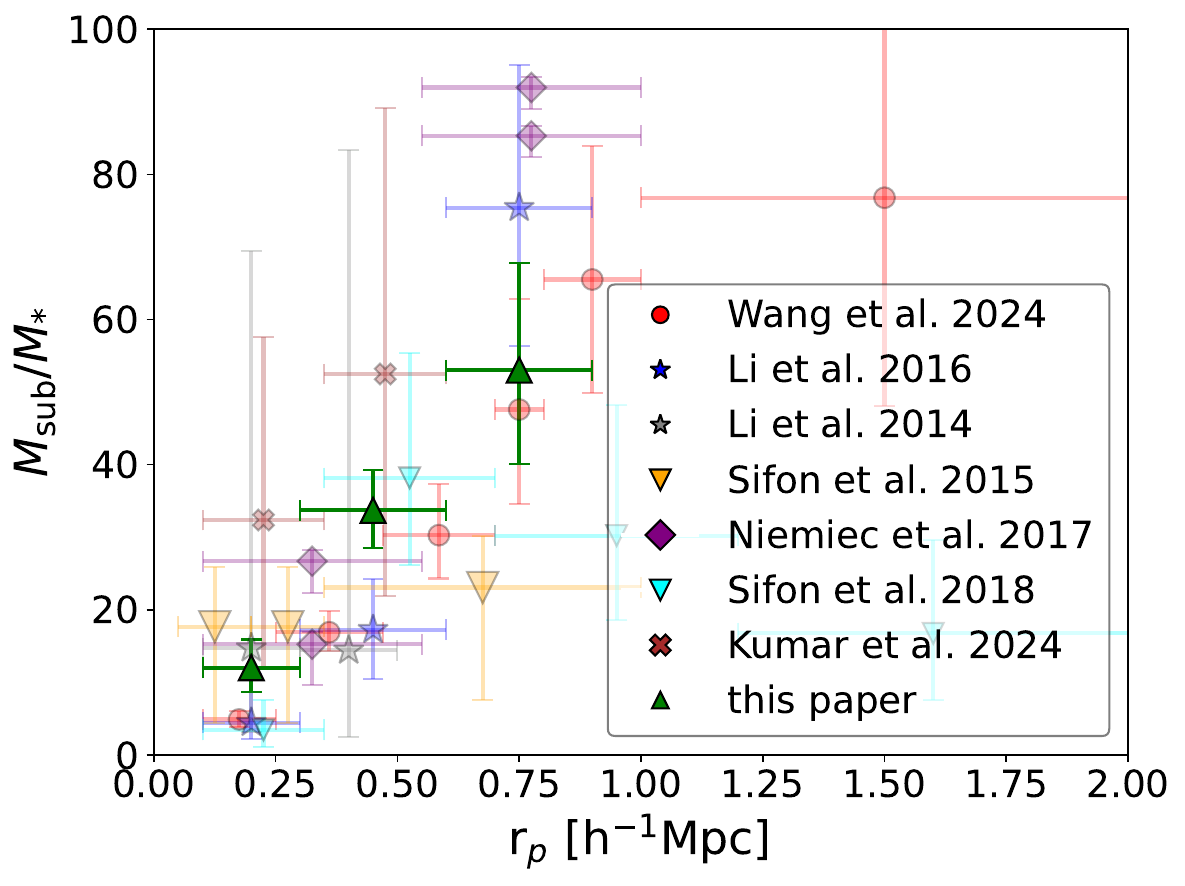}
    \caption{Subhalo mass to stellar mass ratio inferred from this study (green triangles) compared to previous results cited in the legend. The cluster-centric distance is in the same units and cosmology as adopted in those works. Note that \protect\cite{kumar2024} binned lenses by luminosity, and in this figure we select the (10.1-11.0]/log($h^{-2}L_{\odot}$) bin. Values of \protect\cite{wang24} are taken without stellar mass binning. Referenced works study satellite populations with different stellar masses and mass definitions than our analysis. This figure aims to illustrate consistent trends with radius, rather than matching amplitudes. }
    \label{fig:HSMR}
\end{figure}

In Figure \ref{fig:HSMR} we compare our posterior results for the HSMR with the results from previous works. We follow the observed trend of decreasing subhalo mass as the satellite is drawn closer to the cluster centre. Despite a noticeable scatter in halo and stellar masses, we confirm tidal stripping of dark matter subhalos for the redMaPPer clusters. All our posterior values are reported in Table \ref{tab:param_fits}. Our test approach of utilizing only satellites within common halos did not affect the reported trend in HSMR.

We also compared our results for BCG mis-centring, represented by Equation \ref{eqn:rayleigh} and the $\sigma$ parameter that defines the Rayleigh distribution, to those of \cite{zhang2019dark}. In their study, the authors model the centring performance of the redMaPPer cluster finding algorithm using archival X-ray observations. They find that the mis-centring offset follows a Gamma distribution, but they also provide a parametrized Rayleigh distribution (see Table 2 of \cite{zhang2019dark}). We can compare their estimate of mean miscentring of the entire redMaPPer cluster population, allowing for the fact that in their model 70\% of clusters are well-centred and 30\% have large offsets, with our estimates, applying the richness-dependent scaling $R_{\lambda}=(\lambda/100)^{0.2}h^{-1}$Mpc, where $\lambda$ is the mean richness of our $r_p$ subsamples (see Table \ref{tab:param_fits}). The results of this comparison are presented in Table \ref{tab:mis-center}. The closest cluster-centric bin shows the greatest agreement between the two distributions, whereas the other two bins exhibit substantial divergence. This divergence is correlated with cluster-centric distance and does not appear to be driven by the increase in the mean richness of the halo subsamples. It should be noted that \cite{zhang2019dark} uses a different subsample of redMaPPer clusters with a cut-off in cluster richness and redshift that we do not apply. 
Although a direct comparison between the two studies is not straightforward, the increase in offset distances with cluster-centric bin in our data once again requires further investigation.

\begin{table}
    \centering
    \caption{Comparison of the mean of the Rayleigh distribution that describes the redMaPPer BCG mis-centering, derived from the analysis of \protect\cite{zhang2019dark} and this work.}
    \begin{tabular}{lcc}
         &  \multicolumn{1}{c}{Zhang et al. 2019}&  \multicolumn{1}{c}{This work}\\
         \toprule
         $r_p$ [Mpc] 
&  mean [kpc]&  mean [kpc]\\
\midrule
         0.143--0.43 
&  100.48$^{+22.34}_{-24.82}$&  104.66$^{+16.16}_{-18.64}$\\
         0.43--0.86 
&  101.48$^{+22.56}_{-25.08}$&  166.33$^{+16.61}_{-19.39}$\\
         0.86--1.23 
&  105.29$^{+23.41}_{-26.01}$&  435.77$^{+29.26}_{-37.20}$\\
         \bottomrule
    \end{tabular}
    
    \label{tab:mis-center}
\end{table}

While constructing a subhalo profile, we considered describing it using a truncated NFW profile (tNWF) from \cite{baltz09}, which could constrain the presence of tidal stripping. 
In this profile, $\tau = r_t/ r_s$ is a truncation parameter with $r_t$ a truncation radius. 
By putting $\tau$ as a free parameter, we could correlate it with the cluster-centric distance ($r_p$) of the satellite. However, as found in \cite{li16}, and in our initial model fitting, $\tau$ is poorly constrained at low values and has a flat likelihood when it approaches prior boundaries. No correlation with $r_p$ was found either.

In Appendix \ref{sec:R0R1}, we utilize a different host halo model that was explored in the fitting process. 

\section{Conclusion}
\label{sec:concl}

In this work, we combined redMaPPer SDSS DR8
cluster and member catalogs with the ShapePipe (v1.3) weak-lensing sources catalog from UNIONS to compute the galaxy-galaxy lensing signal as excess surface density. Computations were performed using \texttt{dsigma}, and the results display similar signal morphology as previous research. By assigning halo photometric redshift to lensing galaxies with unknown spectroscopic redshift, we greatly increased the number of lens-source pairs. Combined with a big sky overlap in the Northern hemisphere between lenses and sources catalog, this produced a signal with very low noise. 

The simple and conservative model that consists of halo, subhalo, and stellar terms was applied in the fitting process. This model was previously used in \cite{wang24}, \cite{li16}, \cite{kumar2024}, \cite{niemiec17}. 
Following the idea proposed in \cite{kumar2024}, we sampled from a Rayleigh distribution of BCG offset distance. This sampling allowed us to correctly model the signal at the minima by tweaking the scale parameter of the Rayleigh distribution. Scale parameter $\sigma$ is also the mode of the distribution and contains a physical meaning of the most common offset distance between redMaPPer BCG and ``true'' BCG. Results from our fitting routine showed that $\sigma$ correlates with cluster-centric bins, meaning that, on average, the uncertainty in BCG distance is larger for the host halo sample of satellites further away. This also could be explained by different host halo samples for each lensing satellite bin, as we show that the average host halo richness is also correlated with cluster-centric distance bins.

Another fitting routine was applied, where the offset host halo term was a linear combination of an offset halo term computed using distances to the most likely central galaxy and to the second likely central galaxy. Here we dropped the $\sigma$ parameter and replaced it with 2 linear coefficients for 2 terms. Unlike our final model, which struggles to explain certain regions of the data by underestimating amplitude or not following a slope, this one provides a better fit overall. It may not be physical, but it shows that offset halo modeling and the uncertainty in projected cluster-satellite distances would benefit from a more constructive approach. 

 We believe that some other assumptions need to be made that can replicate halo-subhalo interactions in regions that were poorly explained by our model. The simulations in this work do not include substructures of the cluster halo, which may play a bigger role. Perhaps a simple NFW profile for a subhalo and a halo is not enough. Radial symmetry of our model could be broken with a more realistic consideration of the effect on the observed signal by neighbouring satellites. We believe that a simple approach of Monte-Carlo simulation could be used to create an offset halo profile that considers all satellites' positions for a single halo simultaneously. An addition of a second halo term is also worth considering, as the uncertainty in BCG distance that is favored by data could be explained by it. 
 
 It is also worth mentioning that there was no investigation into the potential bias that could be imposed by our redshift substitution. With that said, we anticipate more substantial redshift estimations for lensing galaxies in the near future. An updated UNIONS catalog will improve errors even further, meaning a more detailed model is needed to simulate the observed lensing signal in galaxy clusters. Nevertheless, our current work allowed to constrain subhalo masses and further support the theory of tidal stripping.

\section*{Acknowledgements}

We want to express gratitude to the University of Waterloo and their initiative  \textit{Summer Internship Program for students at risk from Ukraine} that made this work possible. We also want to thank the Mitacs organization that funded this project via the Mitacs Globalink Research Award. We are
grateful for the computing resources of the Digital Research Alliance
of Canada that was used in this work.

MJH and LVW acknowledge support from their respective NSERC Discovery grants. HH is supported by a DFG Heisenberg grant (Hi 1495/5-1), the DFG Collaborative Research Center SFB1491, an ERC Consolidator Grant (No. 770935), and the DLR project 50QE2305. 

We are honored and grateful for the opportunity to observe the
Universe from Maunakea and Haleakala, which both have cultural,
historical and natural significance in Hawaii. This work is based on
data obtained as part of the Canada-France Imaging Survey, using
observations obtained with MegaPrime/MegaCam, a joint project
of the Canada-France-Hawaii Telescope (CFHT) and CEA Saclay,
on the CFHT, which is operated by the National Research Council
of Canada, the Institut National des Science de l’Univers of the
Centre National de la Recherche Scientifique of France, and the
University of Hawaii. This research is based in part on data collected
at Subaru Telescope, which is operated by the National Astronomical
Observatory of Japan. Pan-STARRS is a project of the Institute for
Astronomy of the University of Hawaii, and is supported by the NASA
SSO Near Earth Observation Program under grants 80NSSC18K0971,
NNX14AM74G, NNX12AR65G, NNX13AQ47G, NNX08AR22G,
80NSSC21K1572, and by the State of Hawaii.

\section*{Data availability}

The satellite galaxy and cluster sample used in this analysis is
constructed from publicly available redMaPPer cluster/member catalog on VizieR service [DOI:\href{https://doi.org/10.26093/cds/vizier.22240001}{10.26093/cds/vizier.22240001}]. The satellite galaxy stellar mass data was taken from publicly available DESI project catalog on VizieR service [J/ApJS/242/8].
The raw and processed UNIONS data are currently
available to members of the Canadian, French, Japanese, and
Pan-STARRS communities. All UNIONS data will be publicly
available to the international community at the end of the proprietary period.



\bibliographystyle{mnras}
\bibliography{refs} 



\appendix
\section{redshift substitution}
\label{sec:redshift}

In order to achieve more lens-source pairs, while not sacrificing the accuracy of the measured lensing signal, we need to filter satellite galaxies from the redMaPPer catalog. The first and most reasonable way is to select by the satellite's cluster membership probability, which is estimated according to their photometric
redshift, color, and cluster-centric distance. The lensing model we study is very sensitive to contamination by field galaxies, i.e., galaxies that are not part of the cluster. 
This is significant because if dark matter subhalos around satellites are predicted to be stripped, then field galaxies, which do not experience tidal interactions will contaminate the interpretation of the signal \citep{sifon15}. To avoid that, we select galaxies with a membership probability of $P_{\rm mem}>0.8$.  Still, many likely members of clusters lack accurate spectroscopic redshift measurements, which are essential for lensing signal estimation. In this work, we assigned the cluster's photometric redshift to satellites with unknown spectroscopic redshift to increase the total number of lens-source pairs. Photometric redshifts in the redMaPPer cluster catalog are constrained by fitting all possible member galaxies simultaneously to the red-sequence color function. We assume that the cluster members lie on the same plane in the sky as the cluster centre, flattening the forward or backward satellite position in this approximation. The resulting number of lenses filtered for signal computation and offset halo modeling using different selection criteria is summarized in Table \ref{lenstable}. 

\begin{figure}
    \centering
    \includegraphics[width=0.46\textwidth]{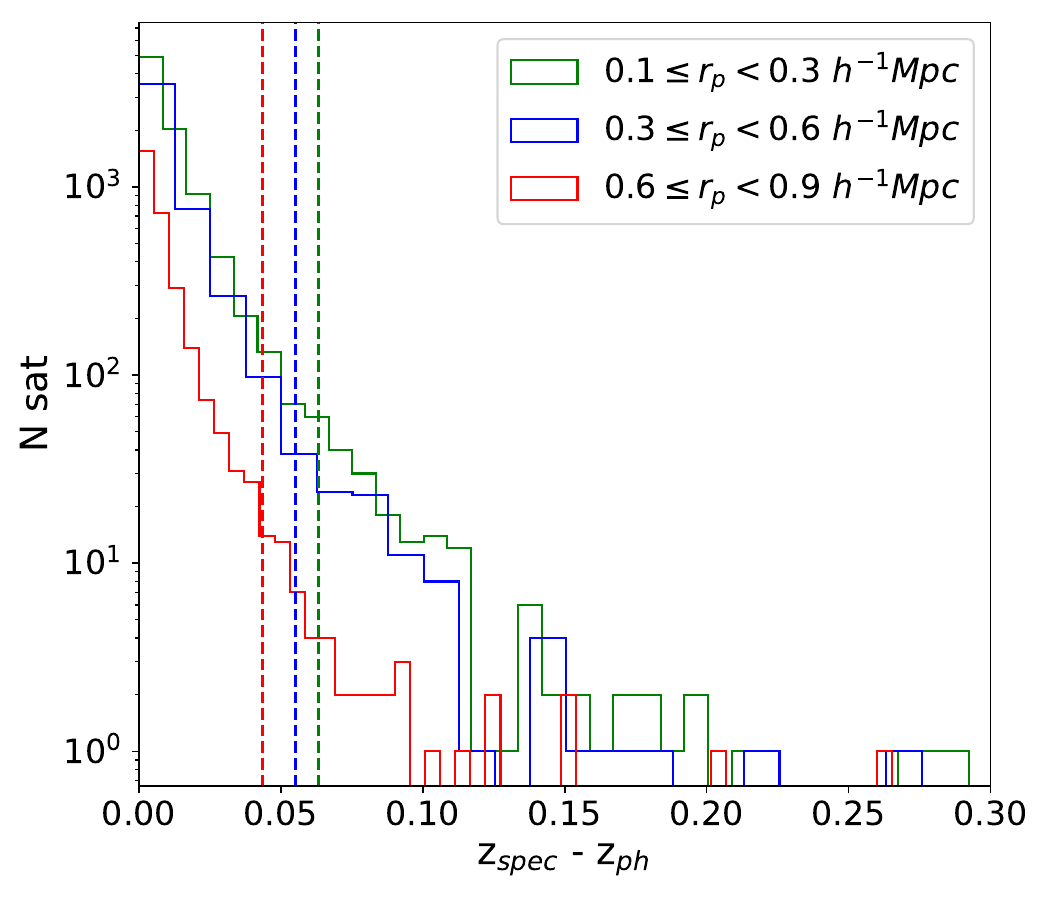}
    \caption{Histograms of differences between members' known spectroscopic redshift and cluster's photometric redshift for each cluster-centric bin. Dashed lines indicate the largest redshift difference for 98\% of the galaxies in each bin. }
    \label{fig:zdiffzspec}
\end{figure}

In this section, we show our statistical analysis in order to justify redshift substitution. Relative to large-scale structure and the redshift of background galaxies, the redshift offset between the cluster centre and its satellite galaxies is insignificant. A major discrepancy between photometric and spectroscopic redshift would arise from Lyman break estimation, given the spectral resolution of the instrument. We compute a difference between known $z_{\text{spec}}$ of the satellite and corresponding halo's $z_{\text{ph}}$ in Figure \ref{fig:zdiffzspec} for a Pmem$>$0.8 selection. In the $0.6\leq R_p<0.9$ Mpc $r_p$ range, we have 329451 lenses in total, but only around 6\% have spectroscopic redshifts in the catalog. In this sample, 85\% of the lensing galaxies have a redshift difference lower than 0.02. With this result, we decided to safely assign cluster $z_{\text{ph}}$ to lenses with unknown redshift.

\begin{table}
\caption{Different cluster member selection filters with the corresponding number of satellites. The first filter includes only satellites with spectroscopic redshifts. The second selects members with a probability $>80\%$ of being cluster satellites. The next row contains only lenses that follow the previous filter and are all in the same clusters. The final row combines both membership probability and redshift criteria, resulting in a smaller sample. Note that satellite galaxies with spectroscopic redshift are not guaranteed to have a P$_{\rm mem}$ of 1 or 0 because it is estimated from the red sequence parameters \citep{rykoff14}.}
\footnotesize
\setlength{\tabcolsep}{4pt} 
\centering
\begin{tabular}{@{}l@{}ccc}
\hline
\begin{tabular}[c]{@{}l@{}}Selection\\filter\end{tabular} 
  & \begin{tabular}[c]{@{}c@{}}\scriptsize 0.1 $< R_p$ \\ \scriptsize$\leq$ 0.3 \\ \scriptsize $h^{-1}$Mpc \end{tabular}
  & \begin{tabular}[c]{@{}c@{}}\scriptsize 0.3 $< R_p$ \\ \scriptsize$\leq$ 0.6 \\ \scriptsize $h^{-1}$Mpc \end{tabular}
  & \begin{tabular}[c]{@{}c@{}}\scriptsize 0.6 $< R_p$ \\ \scriptsize$\leq$ 0.9 \\ \scriptsize $h^{-1}$Mpc \end{tabular} \\ \hline
$\exists\, z_{\mathrm{spec}}$         
  & 11910   & 19443   & 16979   \\ 
Pmem $>$ 0.8                          
  & 204000  & 106586  & 18865   \\ 
 Pmem $>$ 0.8 $\land$ same halos& 60936& 54934&18799\\
 Pmem $>$ 0.8  $\land$$\exists\, z_{\mathrm{spec}}$& 8904    & 8514    & 2950    \\ \hline
\end{tabular}
\label{lenstable}
\end{table}

\section{Host halo admixture model}
\label{sec:R0R1}

In this model, the offset halo term is a linear combination of 2 precomputed offset halos. They are based on distances to the first and second Brightest Central Galaxy (BCG1 and BCG2) in the \cite{rykoff16} redMaPPer catalog. While the selection of satellites is still based on their cluster-centric distances, the offset distance in the Monte-Carlo simulation is set as the distance to BCG1 and BCG2. This results in smearing of the average minima by galaxies that are beyond the selection distance bin. Physically, this method can be explained by taking into account uncertainty in cluster-centric distance estimation of the member galaxies, or the presence of a second halo term.

\begin{figure}
    \centering
    \includegraphics[width=0.46\textwidth]{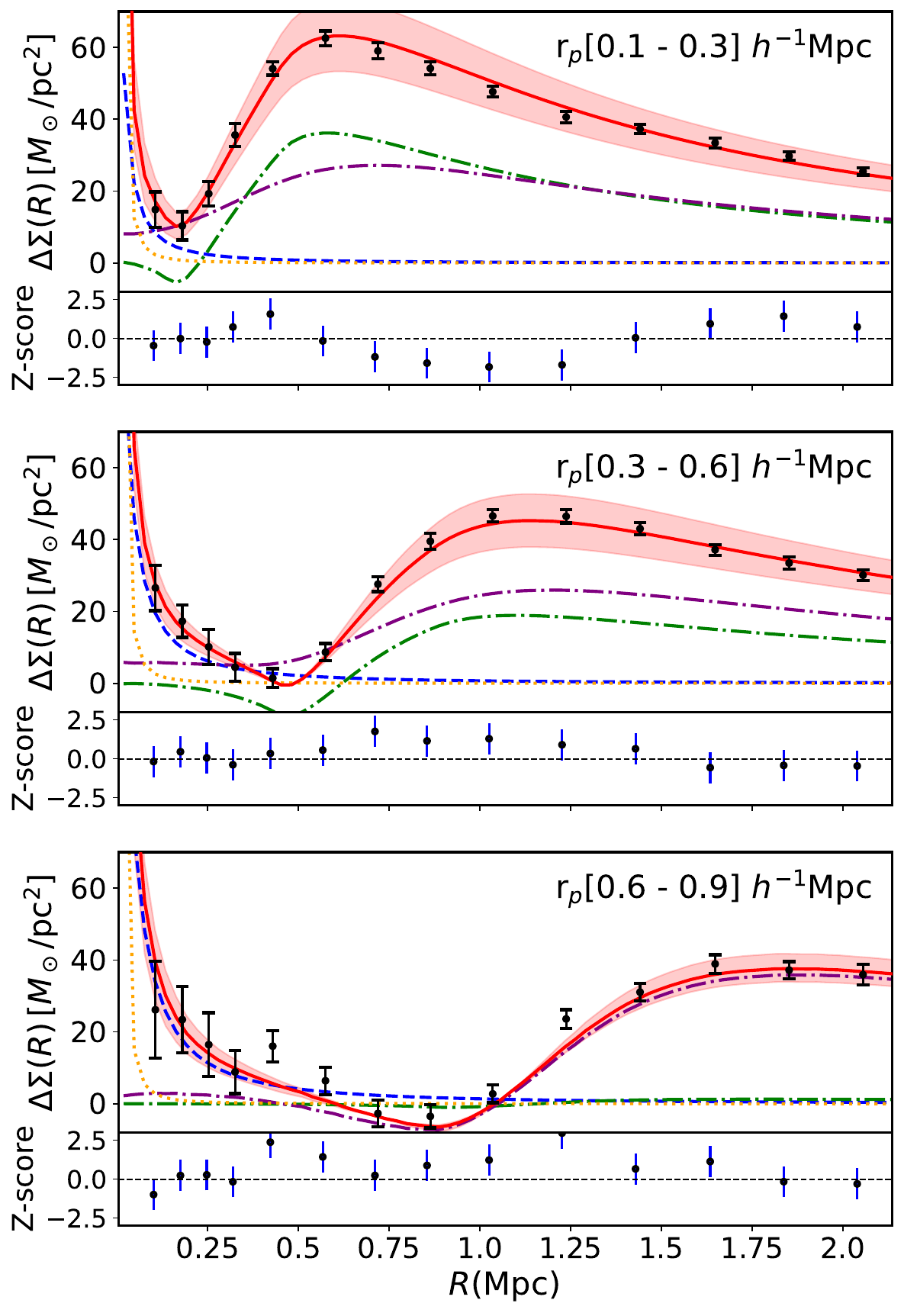}
    \caption{Satellite-centric lensing signal calculated with \texttt{dsigma} and best fit result for each cluster-centric $r_p$ bin. The red solid line is the combined signal that is a sum of 4 terms in our auxiliary model: stellar (orange dashed), subhalo (blue dashed), and 2 offset halo models for the first and second likeliest BCG (green and purple dashed lines). Free parameters are the subhalo mass for subhalo profile estimation, scale factor A, and scale factor B for 2 offset halo profiles. The translucent red area shows the range between the 16th and 84th percentiles of MCMC samples.  Below each fit plot is the corresponding residual graph in units of standard deviation.}
    \label{fig:BCG12}
\end{figure}

\begin{figure}
    \centering
    \includegraphics[width=0.99\linewidth]{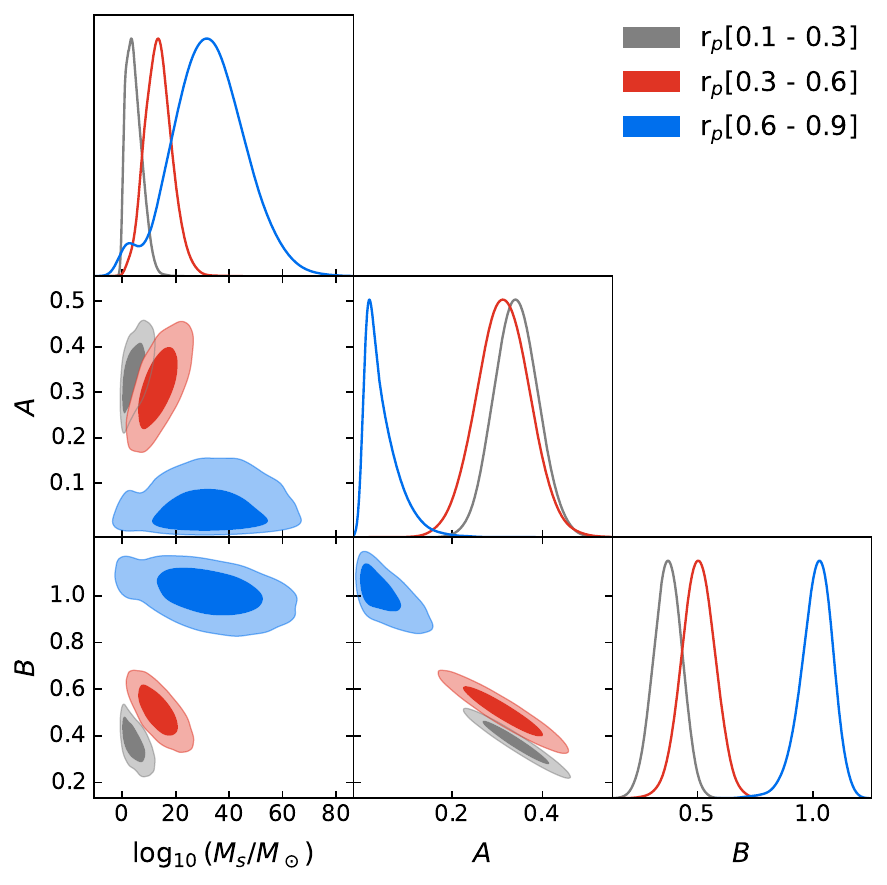}
    \caption{Corner plot of the 2 BCG halo admixture model for each bin. Grey contours for the closest cluster-centric bin, red for intermediate, and blue for furthest. Solid contours around scattered samples indicate 68\% and 95\% of samples, and histograms are marginalized posterior distribution for each fitted parameter (satellite's subhalo mass devided by respective stellar mass (HSMR), offset BCG1 halo scale factor, and offset BCG2 halo scale factor).}
    \label{fig:R0R1_corner}
\end{figure}

For this model, we do not use the Rayleigh distribution of ``true'' BCG. Our fitting parameters are: subhalo mass ($\log_{10}{M_{\text{sub}}}$), BCG1 and BCG2 scaling factors ($A$ and $B$), which are not normalized:

\begin{align}
    \Delta\Sigma_{\rm host} = A \cdot \Delta\Sigma_{\rm bcg1} + B \cdot \Delta\Sigma_{\rm bcg2}
\end{align}

Both priors on $A$ and $B$ follow a uniform distribution between 0 and 1, with the condition that none can go below 0.
The fit accuracy of this model is compared to observed excess surface density in Figure \ref{fig:BCG12}, and the marginalized posterior distribution is shown in Figure \ref{fig:R0R1_corner}. Combined profiles perform better with lower $\chi^2$ values for all bins than our excepted final model. Posterior model parameters are reported in Table \ref{tab:param_fitsR0} along with $\chi^2$. The shape of the observed signal is well replicated by the model at all scales. The ``hump'' at the furthest $r_p$ bin is still unexplained, which confirms the presence of a finer structure that affects the signal around 0.5 Mpc. Interestingly, the data gives more weight to the second BCG profile and favors the complete absence of the first-BCG term for the furthest bin. While the admixture model performs better, it is not physically plausible, and we do not consider it as the final result.

\begin{table}
\centering
\caption{Median parameter values for the  2-halo admixture model with errors based on the 16 and 84 percentiles from the posterior
 distributions. Best-fit estimations of satellite halo mass ($\log\langle M_{sub}\rangle$), BCG halo term rescale factor ($A$),  second BCG halo term rescale factor ($B$), as well as reduced $\chi^2$ number for each cluster-centric bin.}
\begin{tabular}{@{}llllll}
\toprule
$r_p$ [Mpc] & log$\langle M_{sub}\rangle$ & A  &B& $M_{sub}$/$M_{*}$  &$\chi^2_{\text{red.}}$\\ \midrule
0.143-0.43 &$11.54^{+0.27}_{-0.50}$& $0.34^{+0.05}_{-0.05}$&$0.37^{+0.06}_{-0.06}$&$4.01^{+5.51}_{-2.73}$&1.72\\
0.43-0.86 &$12.05^{+0.14}_{-0.20}$& $0.31^{+0.06}_{-0.06}$&$0.50^{+0.07}_{-0.07}$&$13.42^{+5.17}_{-4.92}$&1.17\\
0.86-1.23 &$12.38^{+0.15}_{-0.23}$& $0.03^{+0.05}_{-0.02}$&$1.02^{+0.06}_{-0.07}$&$32.13^{+13.48}_{-13.00}$&1.76\\ \bottomrule
\end{tabular}

\label{tab:param_fitsR0}
\end{table}


\bsp	
\label{lastpage}
\end{document}